\documentclass[slac_one]{revtex4}
\usepackage{graphicx}
\usepackage{fancyhdr}
\usepackage{dcolumn}
\usepackage{graphicx}
\usepackage{bm}
\usepackage{epsfig}
\usepackage{psfrag}
\usepackage{xspace}
\usepackage{multirow}
\usepackage{lscape}
\usepackage{comment}
\usepackage{amssymb}

\newcommand{\ttbar}     {\mbox{$t\bar{t}$}}
\newcommand{\ppbar}     {\mbox{$p\bar{p}$}}

\newcommand{\lplus}     {$\ell$+jets}

\newcommand{\ljets}     {$\ell$+jets}
\newcommand{\met}       {\mbox{$\not\!\!E_T$}}
\newcommand{\pythia}    {\mbox{\textsc{pythia}}}
\newcommand{\alpgen}    {\mbox{\textsc{alpgen}}}
\newcommand{\mcatnlo}    {\mbox{\textsc{mc@nlo}}}
\newcommand{\lumi}      {4.3~$\rm fb^{-1}$}

\newcommand{\toporesult}  {7.70}

\newcommand{\topototal}    {^{+0.79}_{-0.70}} 

\newcommand{\btagresult}  {7.93}

\newcommand{\btagtotal}   {^{+1.04}_{-0.91}} 

\newcommand{\fbmone}      {fb$^{-1}$}

\newcommand{\ltau}      {\ensuremath{\tau \ell}}

\begin{document}

\title{Top Quark Production at the Tevatron} 

%

\author{C. Schwanenberger}
\affiliation{ The University of Manchester, Manchester M13 9PL, United Kingdom }
\author{on behalf of the CDF and D0 Collaborations}

\begin{abstract}
This review gives an overview of most recent measurements of top quark
production cross sections including differential cross sections and
searches for new physics in 
the top quark sector. Datasets corresponding to an integrated
luminosity of up to 5.7~fb$^{-1}$ are presented which were taken at the Tevatron
proton-antiproton collider at a 
center-of-mass energy of $\sqrt{s}=1.96$~TeV at Fermilab.
\end{abstract}

\maketitle

\thispagestyle{fancy}


\section{TOP QUARK PRODUCTION AT THE TEVATRON} 
Top quarks were first
observed via pair production at the Fermilab Tevatron Collider in
1995~\cite{top-obs-1995}. The Tevatron is still the only place where
top quarks can be studied with a very high precision. At the Tevatron, top
quarks are either produced in pairs via the strong interaction or
singly via the electroweak interaction~\cite{singletop-willenbrock}.
In the
framework of the standard model (SM), each top quark is expected to
decay nearly 100\% of the times into a $W$~boson
and a $b$~quark~\cite{top-mass-properties}.
$W$ bosons can decay hadronically into
$q\bar{q}^\prime$ pairs or
leptonically into $e\nu_{e}$, $\mu\nu_{\mu}$ and $\tau\nu_{\tau}$   with
              the $\tau$ in turn decaying into an electron, a muon, or  hadrons,
              and associated neutrinos.

In \ttbar\ production, if both $W$ bosons decay hadronically the final state is called
all-hadronic (or all-jets) channel. If one of the $W$ bosons decays hadronically
while the other one produces
a direct electron or muon  or a secondary electron or muon from
$\tau$ decay, the final state is
referred to as the $\ell$+jets channel.
If both $W$ bosons  decay leptonically, this leads to a dilepton
($\ell\ell$) final
state containing a pair of electrons, a pair of muons, or
an electron and a muon, or a hadronically decaying
tau accompanied either by an electron or a muon (the \ltau\ channel).

\section{TOP QUARK PAIR PRODUCTION}

Exploring the top cross section in different decay channels and using
different assumptions is important because signs of new
physics might appear differently in the various
channels.

\subsection{Top Quark Pair Production Cross Section}

The top quark pair production cross section $\sigma_{t\bar{t}}$ is
known to high accuracy in the
SM~\cite{SMtheory_N,SMtheory_C,SMtheory_A,SMtheory_M,SMtheory_K}. The 
measurement of the top 
quark pair production cross section therefore provides an important
test of calculations in higher order Quantumchromodynamics (QCD)
including soft gluon resummations. Any deviation from
the SM prediction of the measured \ttbar\ cross section could either
be a hint for new physics in top quark pair production or in top quark decays.
For example, an exotic decay of
the top quark, such as the decay into a charged Higgs boson and a $b$
quark ($t\rightarrow H^{+}b$) would lead to deviations
of the measured $\sigma_{t\bar{t}}$ in
individual final states compared to the SM prediction.

Usually there are two different techniques to enhance the top pair signal over
the background. The ``topological'' method explores \ttbar\ event
kinematics to distinguish \ttbar\ signal from background. Because of
the large mass of the top quark, \ttbar\  
events are more energetic, central, and isotropic compared with the
dominant backgrounds such as $W$+jets and QCD multijet events, whose
kinematics are more influenced by the boost from the momentum
distribution of the colliding partons. 
The ``counting'' method utilizes the identification of jets originated from $b$ quarks
($b$-tagging). The first method does not rely on the assumption that
a top quark decays into a $b$-quark as opposed to the second one. Thus they
are sensitive to different systematic uncertainties.

\begin{figure}[ht]
\begin{center}
\includegraphics[width=7.0cm]{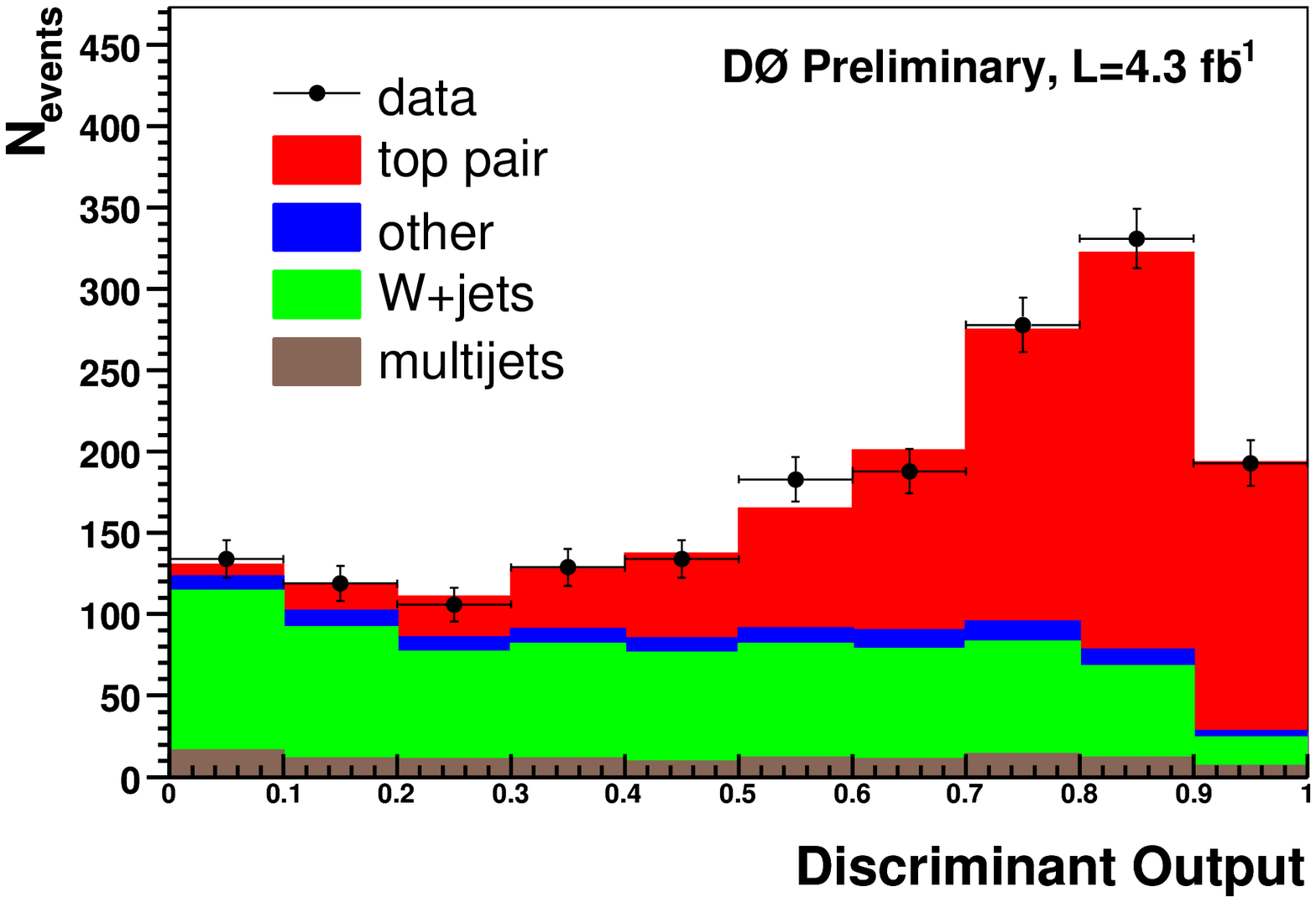}
\includegraphics[width=7.0cm]{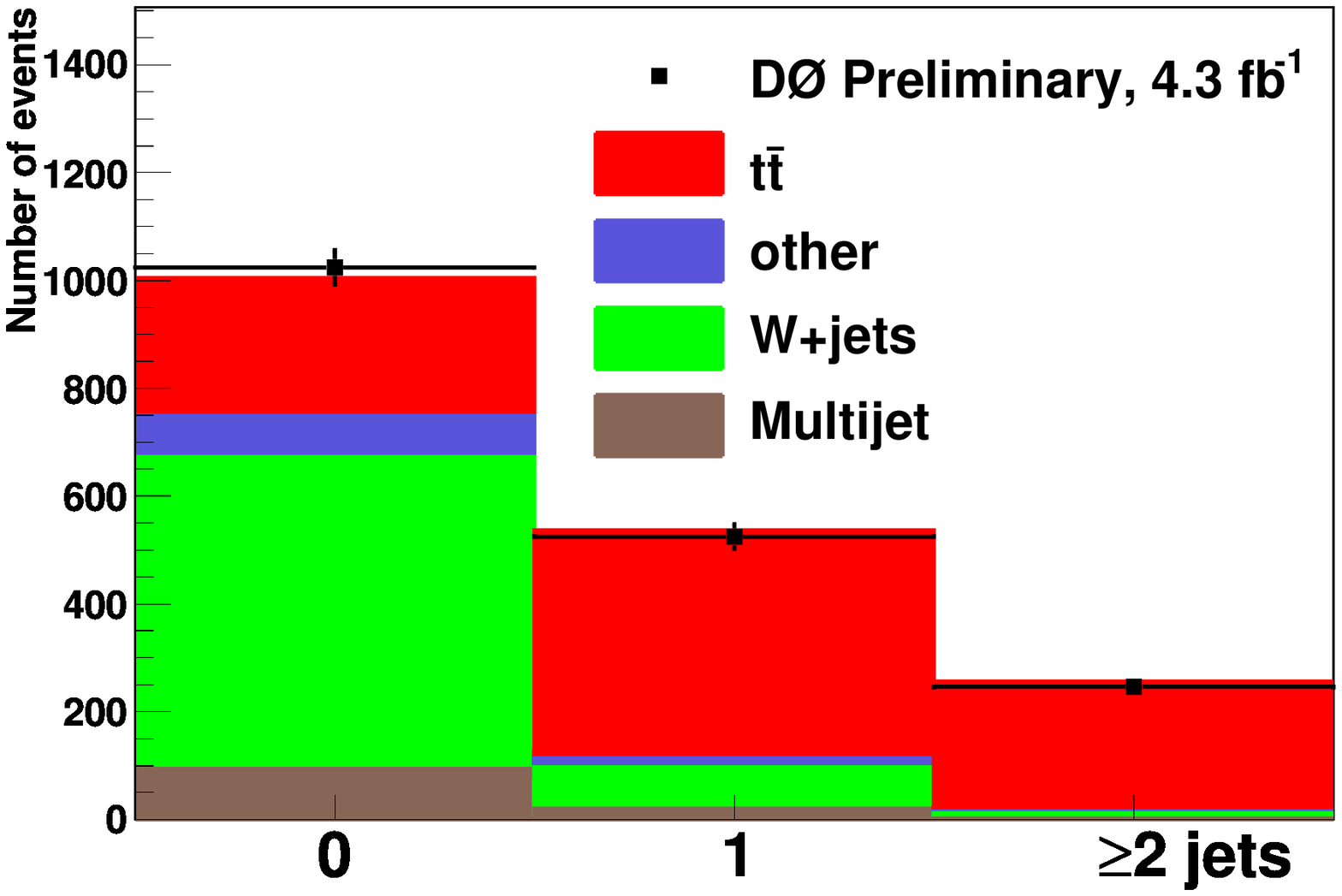}
\end{center}
\caption{Left: output of the BDT discriminant for $\ell$+$\ge4$\,jets,
  showing contributions from \ttbar\ signal using 
a cross section of $7.70$~pb and backgrounds. Right: distribution of events
with $0$, $1$ and $\ge2$ $b$-tagged 
jets for the combined \ljets\ channels, showing contributions from \ttbar\ signal using
a cross section of $7.93$~pb and backgrounds.
\label{fig:d0_xsec} }
\end{figure}

One example for a ``topological'' analysis from the D0 Collaboration~\cite{d0_xsec}
uses an
integrated luminosity of
$4.3$~${\rm fb}^{-1}$
exploring final states with 2, 3 or $\ge 4$ jets, thereby defining twelve
disjoint data sets. 
To distinguish \ttbar\ signal from background, a discriminant is constructed
that exploits differences between kinematic
properties of the \ttbar\ \lplus\ signal and the dominant $W+$jets
background. A multivariant discriminant function is calculated by a random
forest (RF) of boosted decision trees (BDT). 
The output of the BDT discriminant for $\ge 4$ jets is presented in
Fig.~\ref{fig:d0_xsec} (left).
To extract the $t\bar{t}$ cross section one performs a binned maximum likelihood
fit of the discriminant distribution for signal and background to the data.
Systematic uncertainties are accounted for in the maximum likelihood
fit by assigning a nuisance parameter to each independent systematic
variation. Each nuisance parameter 
is modeled as a Gaussian probability
density function with mean at zero and width corresponding
to one standard deviation (SD) of the considered systematic
uncertainty~\cite{nuisance}.
For a top quark mass of 172.5 GeV~\footnote{In the following all
  results are given for a top quark mass of 172.5 GeV, if not
  indicated otherwise. This is within the errors of the current
  Tevatron combination of $173.3 \pm 1.1$~GeV~\cite{mass_wa}.}, one obtains
$\sigma_{t\overline{t}} =
\toporesult \topototal \:{\rm (stat+syst+lumi)}\:{\rm pb}$.

The SM predicts that the top quark decays almost
exclusively into a $W$ boson and a $b$-quark ($t \rightarrow W b$).
Hence besides using just kinematic information, the fraction of $t\bar{t}$ events
in the selected sample can be enhanced using $b$-jet identification.
One example to measure the $t\bar{t}$ cross section in a ``counting''
method is performed by the D0 Collaboration analyzing $4.3$~${\rm
  fb}^{-1}$ of data. Final states with 3 and $\ge 4$ jets are explored.
Each channel is further separated into events with $0$, $1$ and $\ge2$
$b$-tagged jets, thereby obtaining 24 independent sets of data.
Figure~\ref{fig:d0_xsec} (right)
shows a comparison of the distributions in data with $0$, $1$ or $\ge2$ $b$-tagged
jets with the SM $t\bar{t}$ cross section, and the contributions from the
different backgrounds.

An
iterative procedure is used to extract the $t\bar{t}$ cross
section after $b$-tagging. 
The fit of the $t\bar{t}$ cross section ($\sigma_{t\bar{t}}$) to data is
performed using a maximum likelihood fit for the predicted number of
events, which depends on $\sigma_{t\bar{t}}$.
To take into account each channel, the likelihood maximization
procedure multiplies the Poisson probabilities for all channels $j$.
The systematic uncertainties are incorporated in the
fit using nuisance parameters~\cite{nuisance}, each
represented by a Gaussian term. The result is
$\sigma_{t\overline{t}} =
\btagresult \btagtotal \:{\rm (stat+syst+lumi)}\:{\rm pb}$.

\begin{figure}[ht]
\begin{center}
\includegraphics[width=7.0cm]{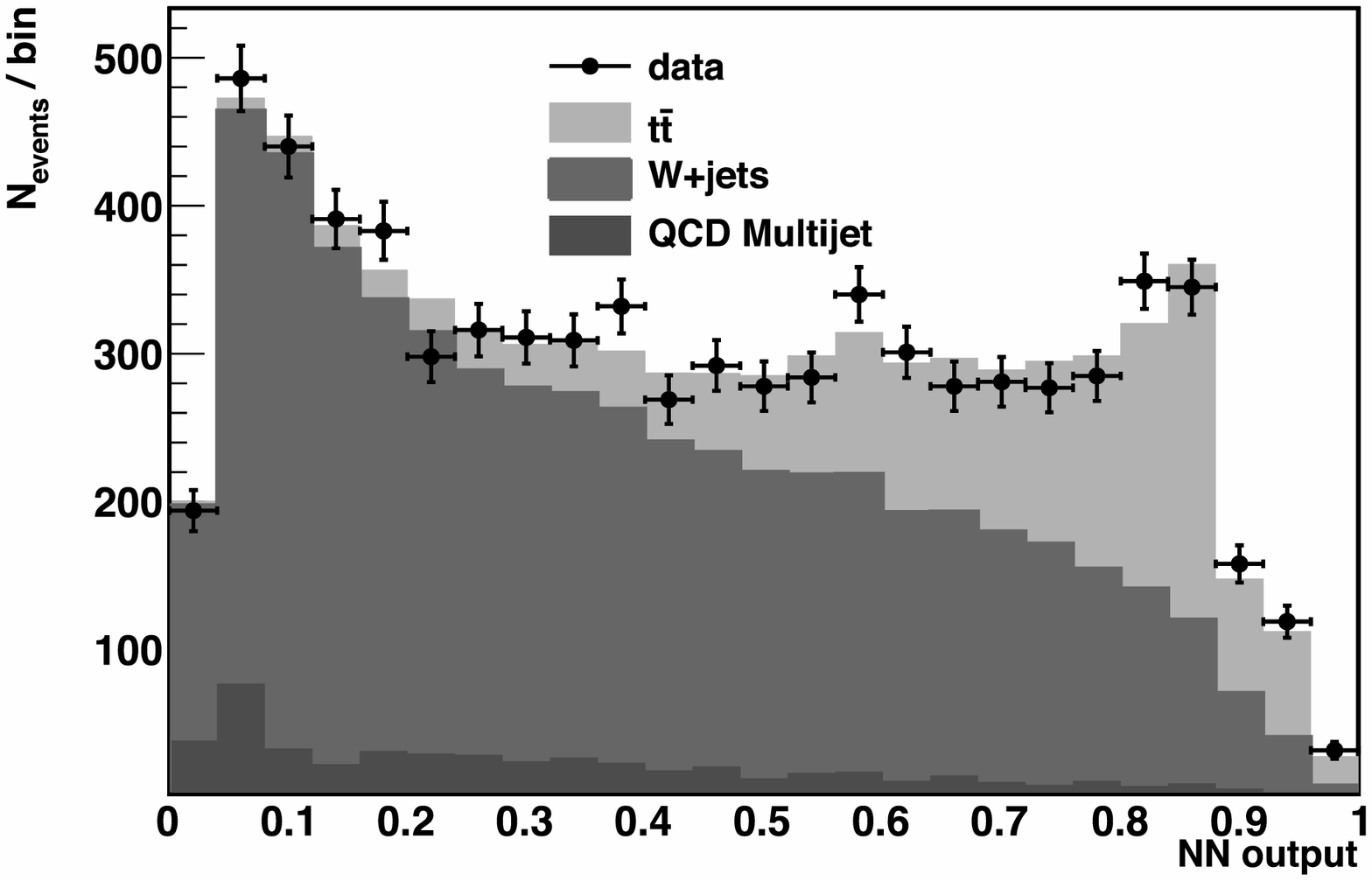}
\includegraphics[width=7.0cm]{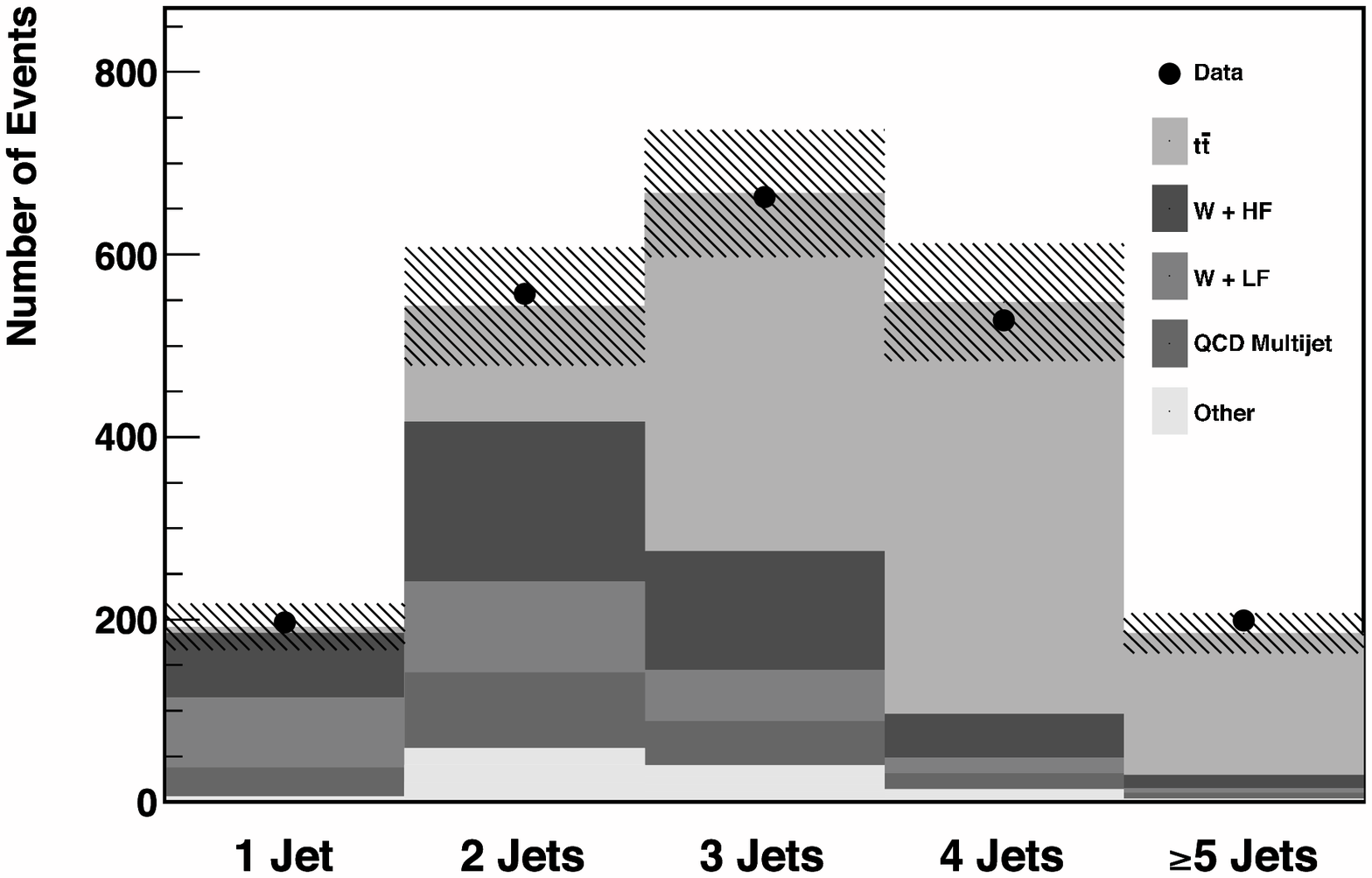}
\end{center}
\caption{Left: output of the ANN discriminant for $\ell$+$\ge3$\,jets,
  showing contributions from \ttbar\ signal using 
a cross section of $7.71$~pb and backgrounds. Right: distribution of events
with $\ge1$ $b$-tagged 
jets in the \ljets\ channel, showing contributions from \ttbar\ signal using
a cross section of $7.22$~pb and backgrounds.
\label{fig:cdf_xsec} }
\end{figure}

The CDF Collaboration, too, performs a ``topological'' measurement of the
top pair production cross section in the \ljets\ channel with an
integrated luminosity of 
4.6~\fbmone\ using an artificial neural network (ANN) technique to discriminate
between top pair production and background
processes~\cite{cdf_xsec}.  The ANN discriminant output is shown in
Fig.~\ref{fig:cdf_xsec} (left). 

The CDF Collaboration also performs a ``counting'' method analyzing
the number of events with at least one $b$-tagged jet. The predicted
number of events for each background process, along with the number of
expected \ttbar\ events at the measured cross section, is shown
compared to data in Fig.~\ref{fig:cdf_xsec} (right).
To measure the \ttbar\ cross section, a likelihood is formed from the
data, the \ttbar\ cross section, and the predicted background for that
cross section.
The result for the ``topological'' method is
$\sigma_{t\overline{t}} =
7.71 \pm 0.37 \:{\rm (stat)} \pm 0.36 \:{\rm (syst)} \pm 0.45 \:{\rm
  (lumi)}\:{\rm pb}$,
the result for the ``counting'' method using $b$-tagging is 
$\sigma_{t\overline{t}} =
7.22 \pm 0.35 \:{\rm (stat)} \pm 0.56 \:{\rm (syst)} \pm 0.44 \:{\rm (lumi)}\:{\rm pb}$.
The largest systematic uncertainties for both experiments come from the measured
luminosity
and the $b$-tag modeling in the
simulation. Because $b$-tagging is not used in the
``topological'' measurement, it is insensitive to the systematic
uncertainty of the $b$-jet tagging measurement.

The dependence on the luminosity measurement and
its associated large systematic uncertainty can be significantly reduced
by computing the ratio of 
the \ttbar\ to $Z/\gamma*$ boson production cross section, measured using the
same triggers and dataset, and then multiply this ratio by the
theoretical $Z/\gamma*$ production cross section~\cite{z_xsec}. In
essence this replaces the 
luminosity uncertainty with the uncertainty on the theoretical $Z/\gamma*$
cross section.
The extracted \ttbar\ cross section is 
$\sigma_{t\overline{t}} = 7.82 \pm 0.38 \:{\rm (stat)} \pm 0.37 \:{\rm
  (syst)} \pm 0.15 \:{\rm (theory)}\:{\rm pb}$  
for the ``topological'' method,
$\sigma_{t\overline{t}} = 7.32 \pm 0.36 \:{\rm (stat)} \pm 0.59 \:{\rm
  (syst)} \pm 0.14 \:{\rm (theory)}\:{\rm pb}$  
for the ``counting'' method with $b$-tagging and 
$\sigma_{t\bar{t}} = 7.70 \pm 0.52 \; \; \mathrm{pb}$ 
for the combination of both.
The uncertainty of $\pm
7\%$ for the ``topological'' measurement is the most precise value
determined by a single measurement to date.

\begin{figure*}[b]
\centering
\includegraphics[width=75mm]{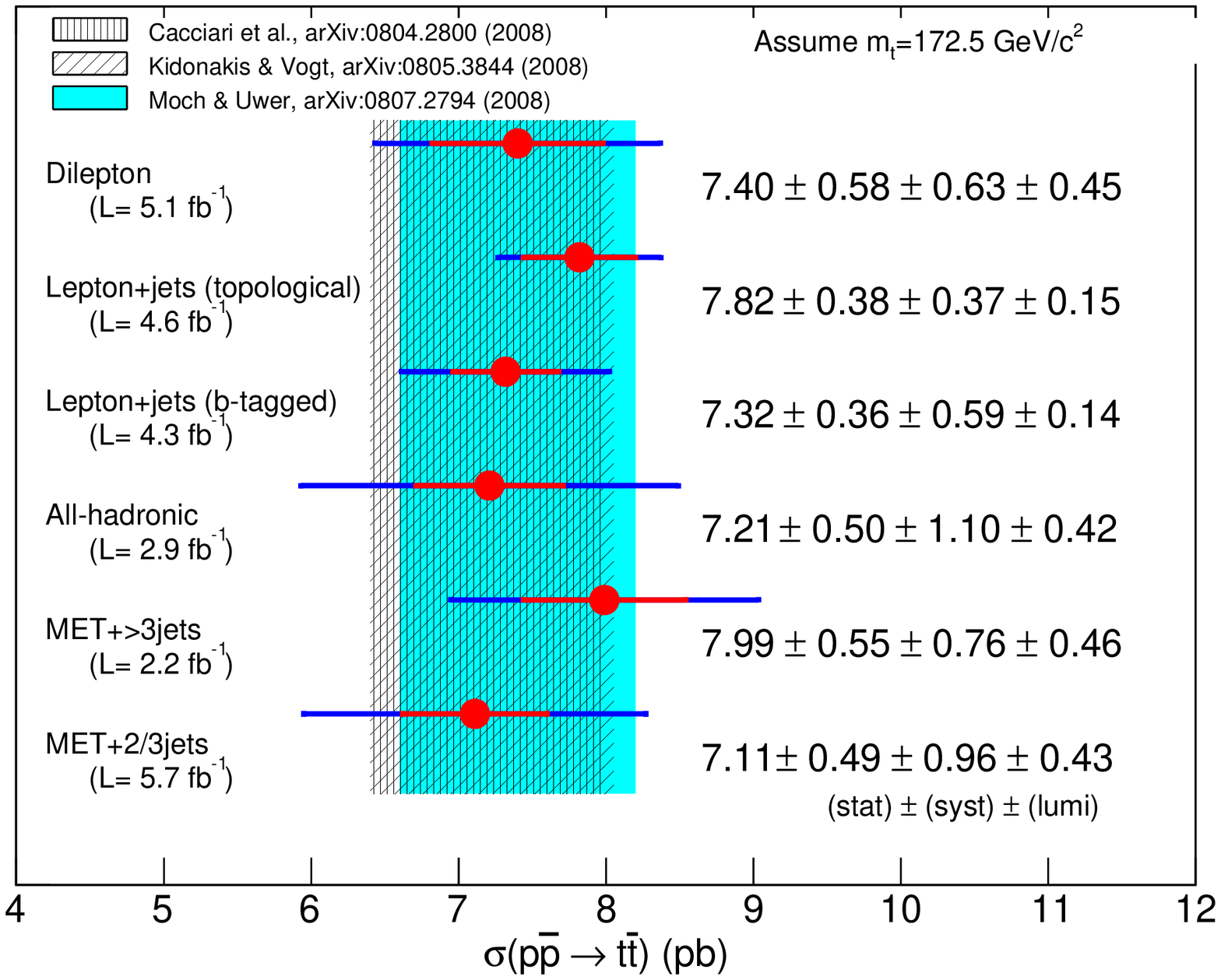}\hspace{1cm}
\includegraphics[width=65mm]{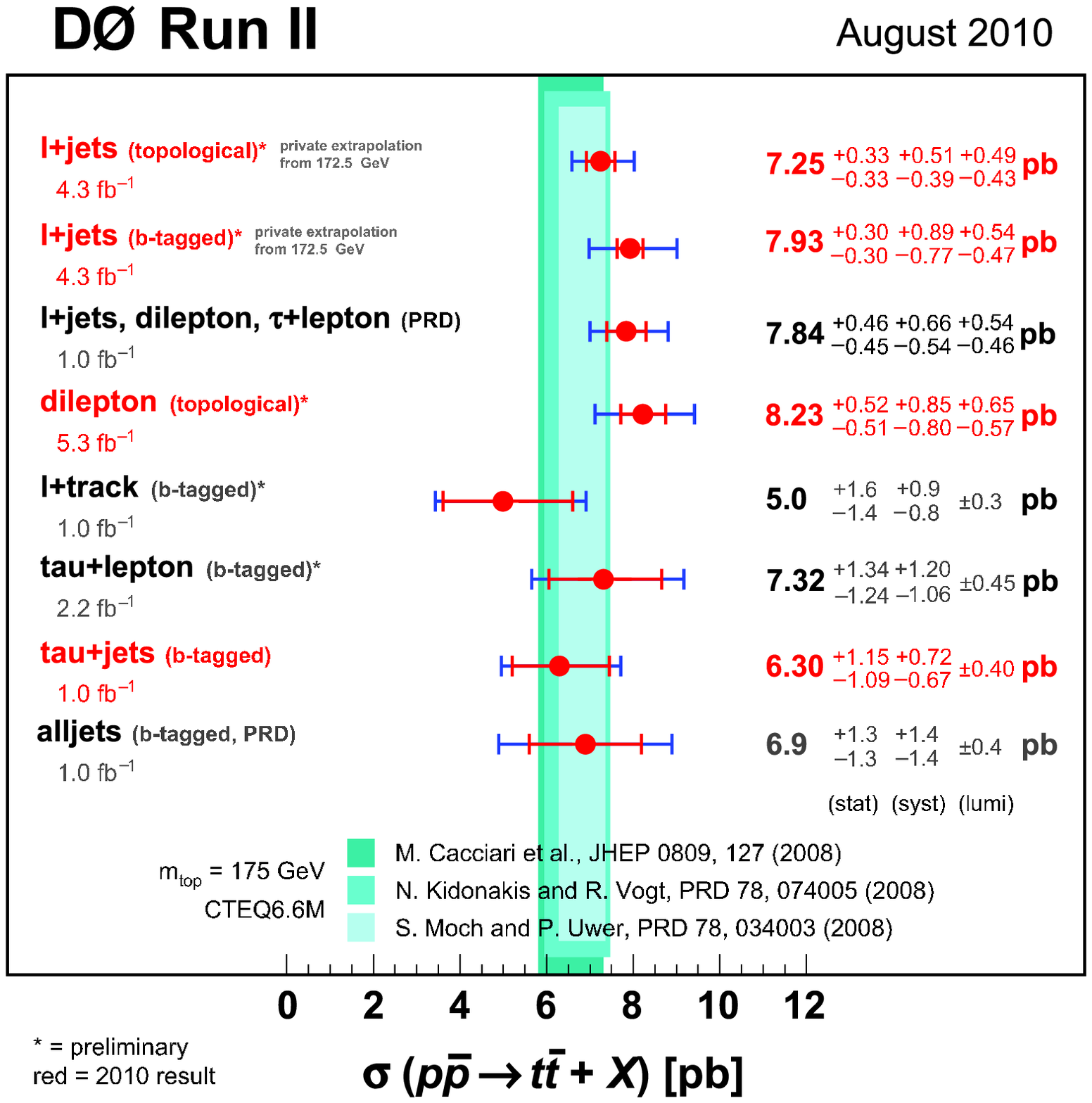}
\caption{Summary of \ttbar\ cross section measurements in different
  decay channels by the CDF (left) and the D0 (right)
  Collaborations. The combination for each experiment is 
displayed, too.
} \label{fig:xsec}
\end{figure*}

Figure~\ref{fig:xsec} summarizes the measurements of the \ttbar\ cross
section in different decay channels performed by the CDF
and D0 Collaborations. All measurements agree with each other and
agree with the SM predictions 
based on the full NLO matrix element~\cite{SMtheory_N} 
including soft-gluon resummation at next-to-leading logarithm (NLL)
accuracy~\cite{SMtheory_C} and at next-to-next-to-leading logarithm (NNLL)
accuracy~\cite{SMtheory_A,SMtheory_K,SMtheory_M}.
The D0 Collaboration has now analyzed all possible decay channels
except final states with two hadronically decaying $\tau$ leptons. In
particular there is a new measurement in the decay channel including a
hadronically decaying $\tau$ lepton and jets~\cite{d0_xsec_taujets}. 
The CDF Collaboration does not perform an explicit identification of
$\tau$ leptons but includes $\tau$ final states in selections of missing
transverse energy and jets~\cite{cdf_xsec_metjets}. This selection
also recovers electrons and muons that were not selected in the
\ljets\ and dilepton channels.
It is also worth mentioning that both collaborations have achieved an
uncertainty of 13\% in measurements of the \ttbar\ cross section
in dilepton final states using a ``topological''
method~\cite{d0_xsec_dilepton} and a ``counting'' method using
$b$-tagging~\cite{cdf_xsec_dilepton}. 
The
CDF Collaboration combines the ``topological'' and ``counting''
measurements of the \ttbar\ cross section in the \ljets\ channel
to the measurements in the
dilepton~\cite{cdf_xsec_dilepton} and
all-hadronic~\cite{cdf_xsec_allhadronic} channels and achieves an
accuracy of $6\%$. The result is
\begin{eqnarray}
\sigma_{t\overline{t}} = 7.50 \pm 0.31 \:{\rm (stat)} \pm 0.34 \:{\rm
  (syst)} \pm 0.15 \:{\rm (theory+lumi\ residual)}\:{\rm pb} \nonumber
\end{eqnarray}
 for a top quark mass of 172.5 GeV.

\subsection{Top Quark Pair + Jet Production}

\begin{figure*}[t]
\centering
\includegraphics[width=75mm]{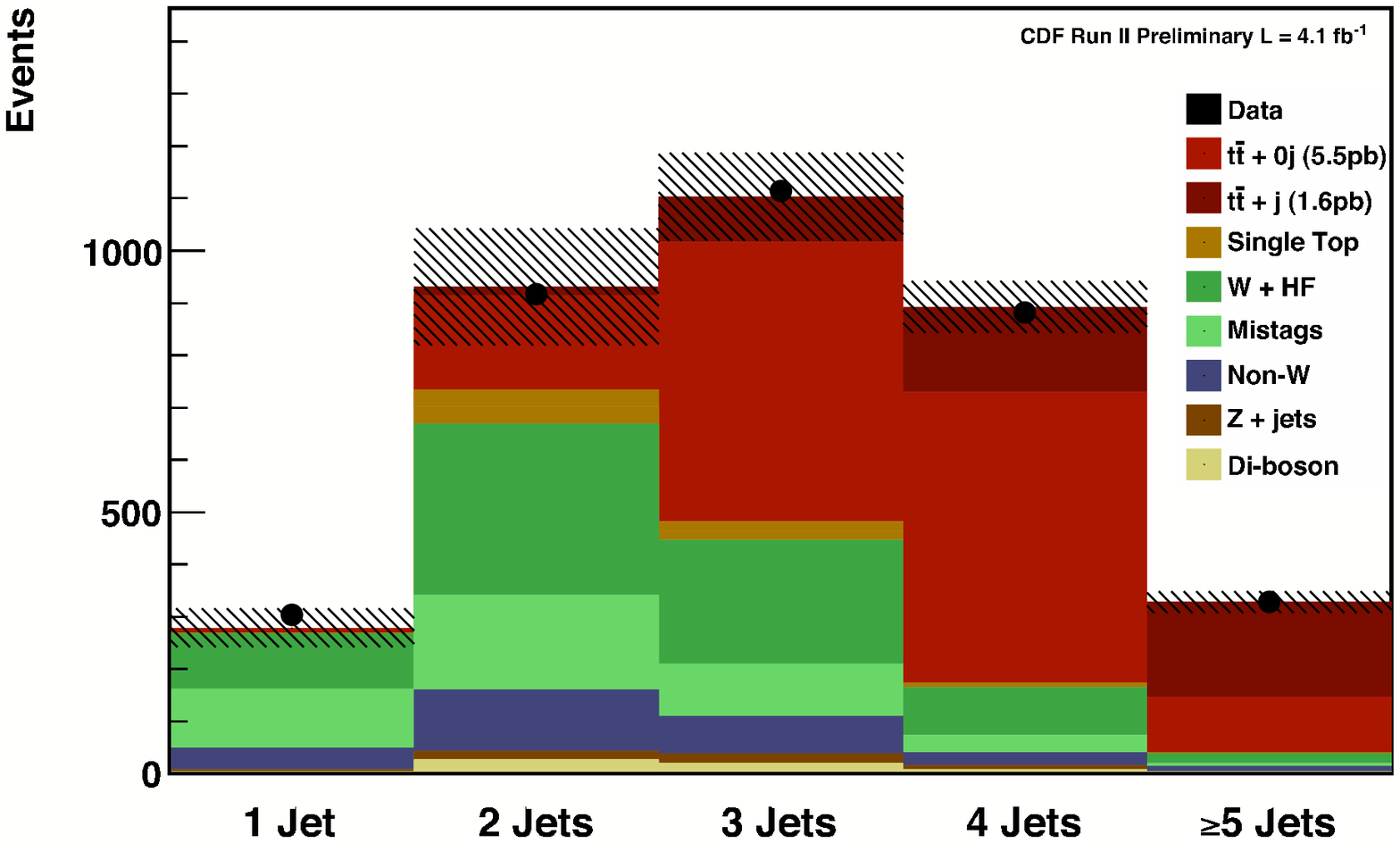}
\includegraphics[width=56mm]{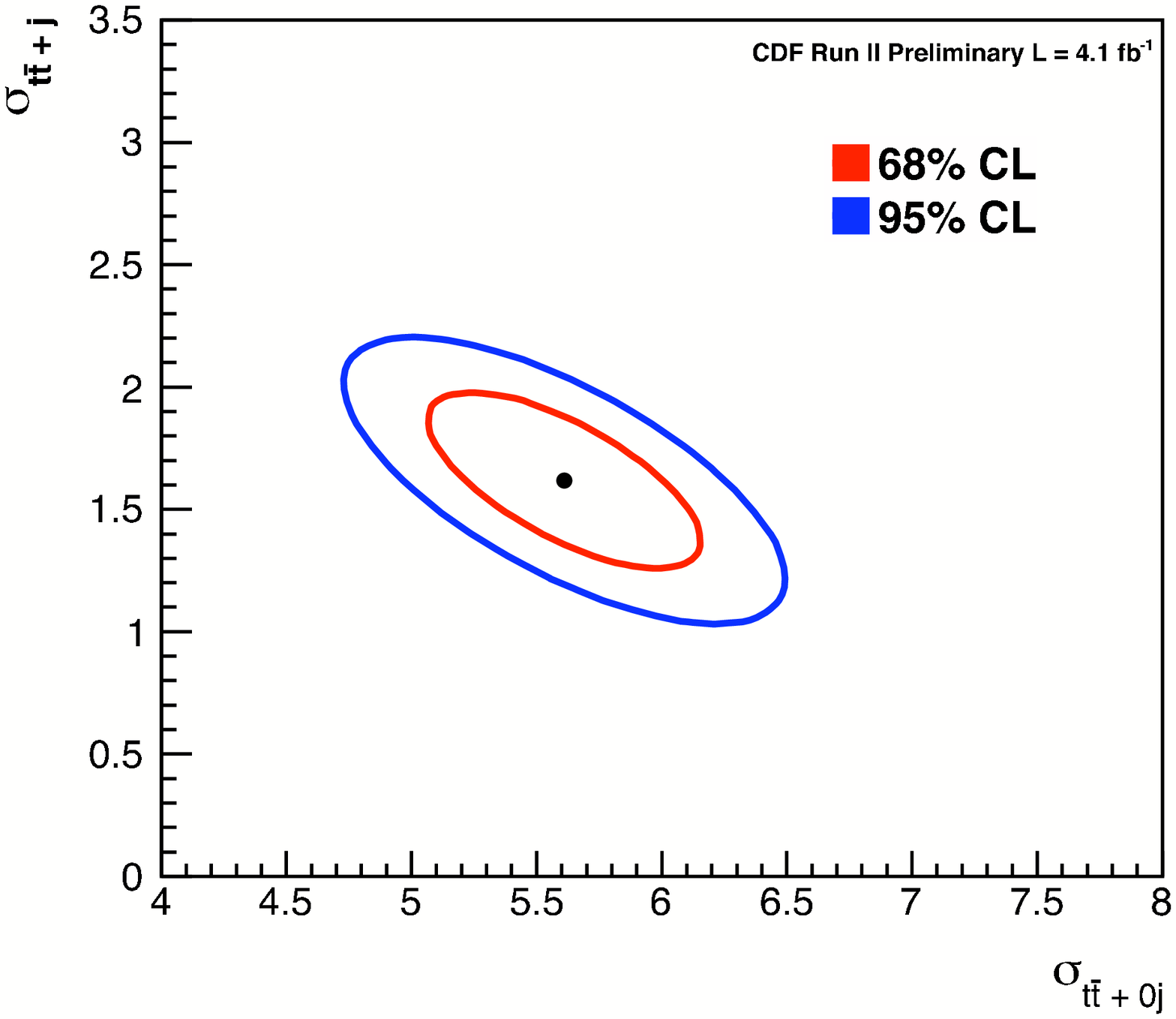}
\caption{Predicted number of events as a function of the jet
  multiplicity (left) and Likelihood curve (right) for the measured \ttbar+0~jet
  and \ttbar+$\ge 1$~jet cross sections.
} \label{fig:cdf_xsec_ttjet}
\end{figure*}

An important test of perturbative QCD, as NLO effects play an
important role in the calculation of the theoretical cross section, is
the production of a \ttbar\ pair associated with an additional hard
jet (\ttbar+jet). This process is interesting in its own right,
because large fractions of the \ttbar\ samples show additional jet
activity and deviations from the SM could signal new physics such as
top-quark compositeness.
The first measurement of the cross section of \ttbar+jet production
has been performed with 4.1~\fbmone\ of collected data by the CDF
Collaboration. The measurement is performed using $b$-tagged events in
the \ljets\ channel.  A data-driven approach is used to predict the
background content. The predicted number of events including the
background, the \ttbar+0~jets and the \ttbar+$\ge 1$~jet contributions is
compared to the data as a function of the jet 
multiplicity in Fig.~\ref{fig:cdf_xsec_ttjet} (left). A 2-dimensional
likelihood is constructed from the data and prediction for events with
three, four, or five jets to simultaneously
measure the \ttbar+jet and the \ttbar\ without jet cross sections. This
is shown in Fig.~\ref{fig:cdf_xsec_ttjet} (right).
The measured result is 
$\sigma_{t\overline{t}+{\rm jet}} = 1.6 \pm 0.2 \:{\rm (stat)} \pm 0.5 \:{\rm
  (syst)}\:{\rm pb}$
which is in agreement with the SM prediction of 
$\sigma_{t\overline{t}+{\rm jet}} = 1.79^{+0.16}_{-0.31}$~pb~\cite{ttj_theo}.

\subsection{Differential Top Quark Pair Production Cross Section}
Measurements of differential cross sections in the \ttbar\ system test
perturbative QCD for heavy-quark production, and can constrain
potential physics beyond the SM. 
The transverse momentum ($p_T$) of top quarks in \ttbar\ events provides a
unique window on heavy-quark production at large momentum scales. The
D0 Collaboration has performed a measurement of the \ttbar\ cross
section as a function of $p_T$ of top quarks for the first time~\cite{d0_xsec_pt}.
Using a data set of 1\fbmone\ the \ljets\ final states with at least
one $b$-tagged jet are explored. 

Objects in the event are associated
through a constrained kinematic fit to the 
$t\bar{t}\to WbW\bar{b} \to \ell\nu b q\bar{q}'\bar{b}$ process.
To obtain a background-subtracted data spectrum, the signal purity
is fitted using signal and background contributions as a function of
 $p_T$, and applied as a smooth multiplicative factor to the data.
The reconstructed $p_T$ spectrum is subsequently corrected for effects
of finite experimental resolution, based on a regularized unfolding
method~\cite{unfolding} using a migration matrix between the reconstructed
and parton $p_T$ derived from simulation. 

\begin{figure*}[t]
\centering
\includegraphics[width=60mm]{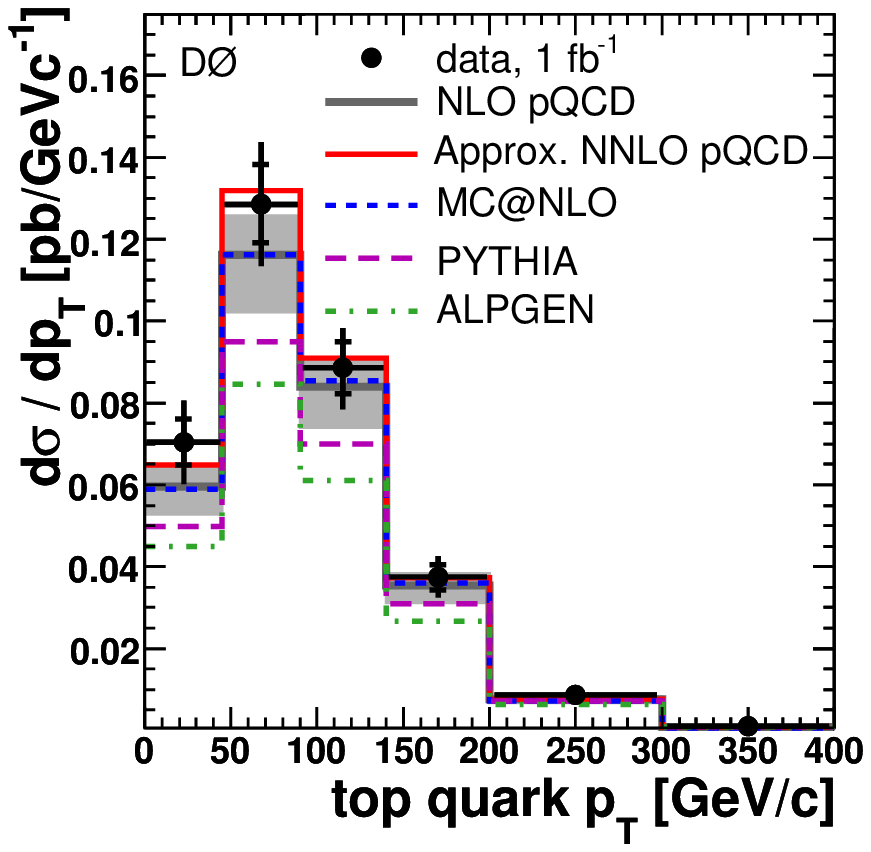}
\includegraphics[width=60mm]{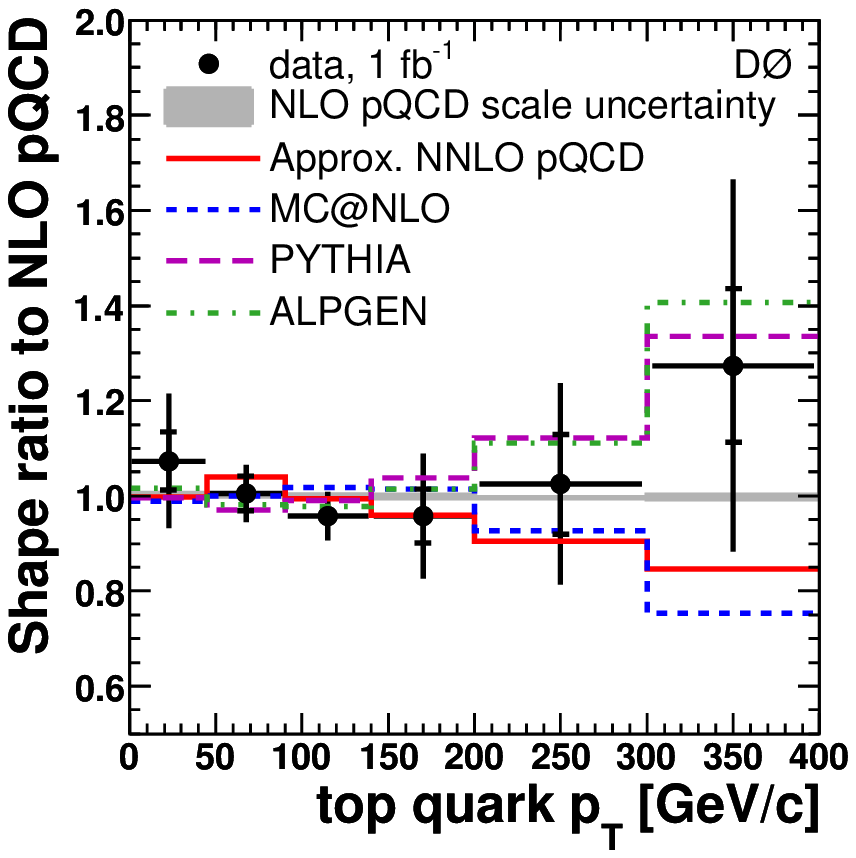}
\caption{
Left: inclusive $d\sigma/dp_T$ for \ttbar\ production (two entries per event for
top and anti-top quark) in data
(points) compared with expectations from NLO QCD (solid lines), from
an approximate NNLO QCD calculation, and for several event
generators (dashed and dot-dashed lines). The gray band represents
the uncertainty on the QCD scale. 
Right: ratio of $(1/\sigma) d\sigma/dp_T$ for top quarks in \ttbar\
production (two entries per event) to the expectation from NLO
QCD. The uncertainty in the scale of QCD is displayed as the gray
band. Also shown are ratios relative to NLO QCD for an approximate
NNLO QCD calculation and of predictions for several event
generators. In both plots inner and outer error bars represent statistical and total
(statistical and systematic added in quadrature) uncertainties,
respectively. 
} \label{fig:d0_xsec_pt}
\end{figure*}

The measured differential cross section as a function of the $p_T$ of the
top quark (using for each event the two measurements obtained from the
leptonic and hadronic top quark decays), $d\sigma / dp_T$ , is shown
in Fig.~\ref{fig:d0_xsec_pt} (left) with the NLO QCD
prediction~\cite{nlo_xsec_pt}.  
Also shown are results
from an approximate next-to-NLO (NNLO) QCD
calculation~\cite{Kidonakis_pt} and from 
the \pythia~\cite{Pythia}, the \alpgen~\cite{Alpgen} (matched with \pythia\
for parton showering~\cite{matching}) and 
the \mcatnlo~\cite{mcnlo} event generators. 
A shape comparison of the ratio of $(1/\sigma) d\sigma/dp_T$ relative to NLO
QCD is shown in Fig.~\ref{fig:d0_xsec_pt} (right). This shows that
results from NLO 
and approximate NNLO QCD calculations and from the 
\mcatnlo\ event generator agree with the normalization and shape of the
measured cross section. Results from \alpgen+\pythia\ and \pythia\
describe the shape of the data distribution, but not its
normalization. 

\begin{figure*}[t]
\centering
\includegraphics[width=60mm]{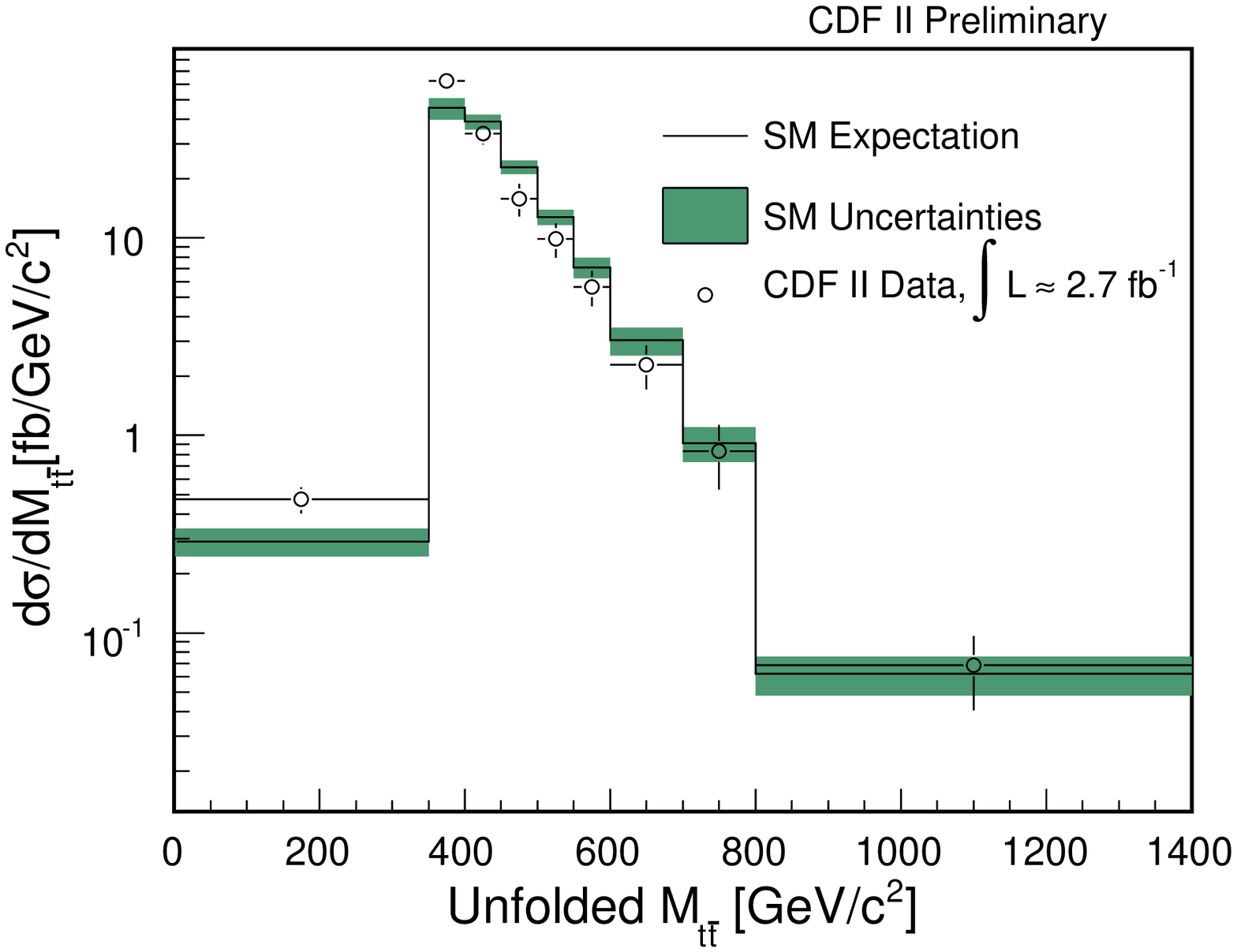}
\includegraphics[width=45mm]{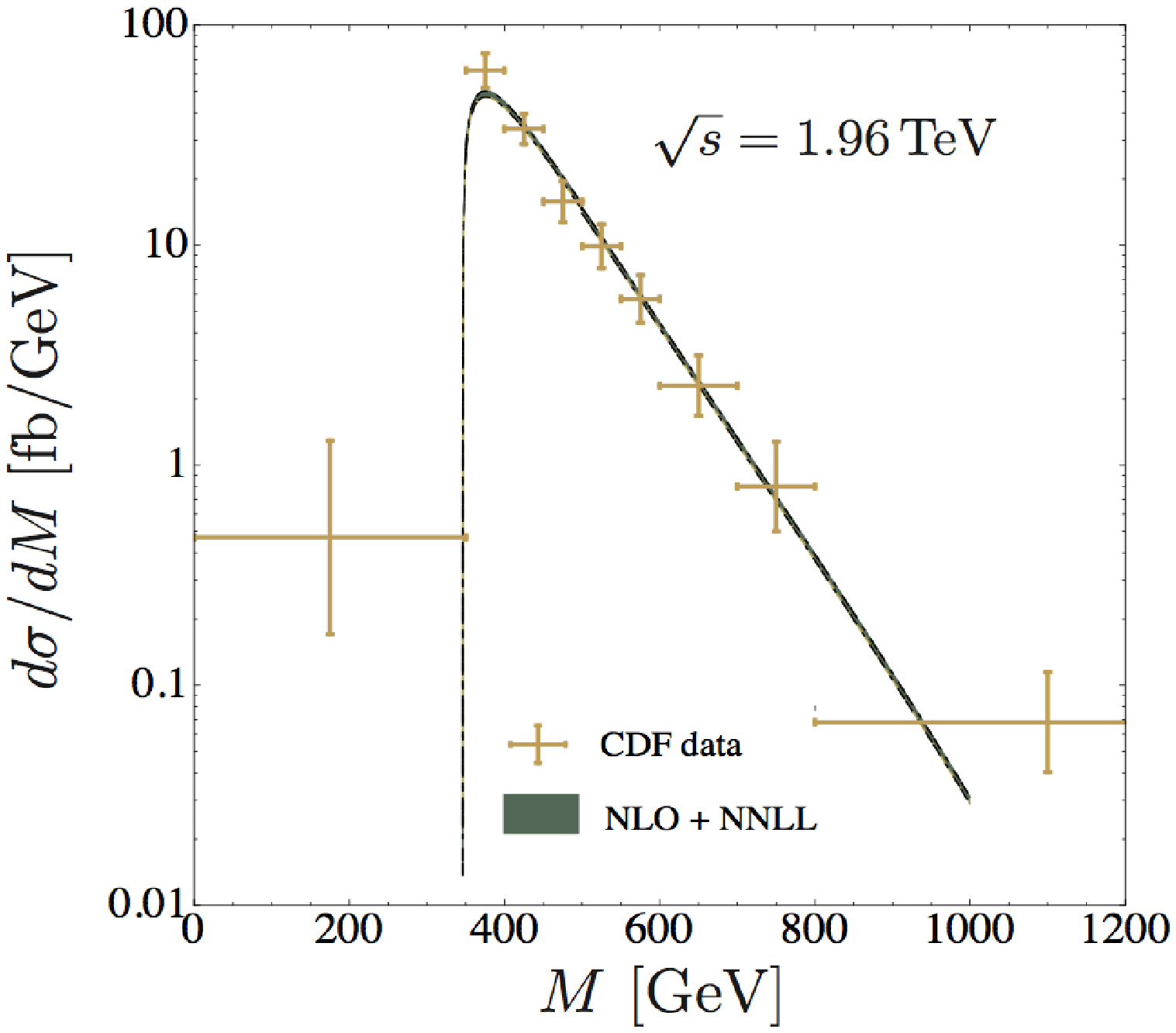}
\includegraphics[width=70mm]{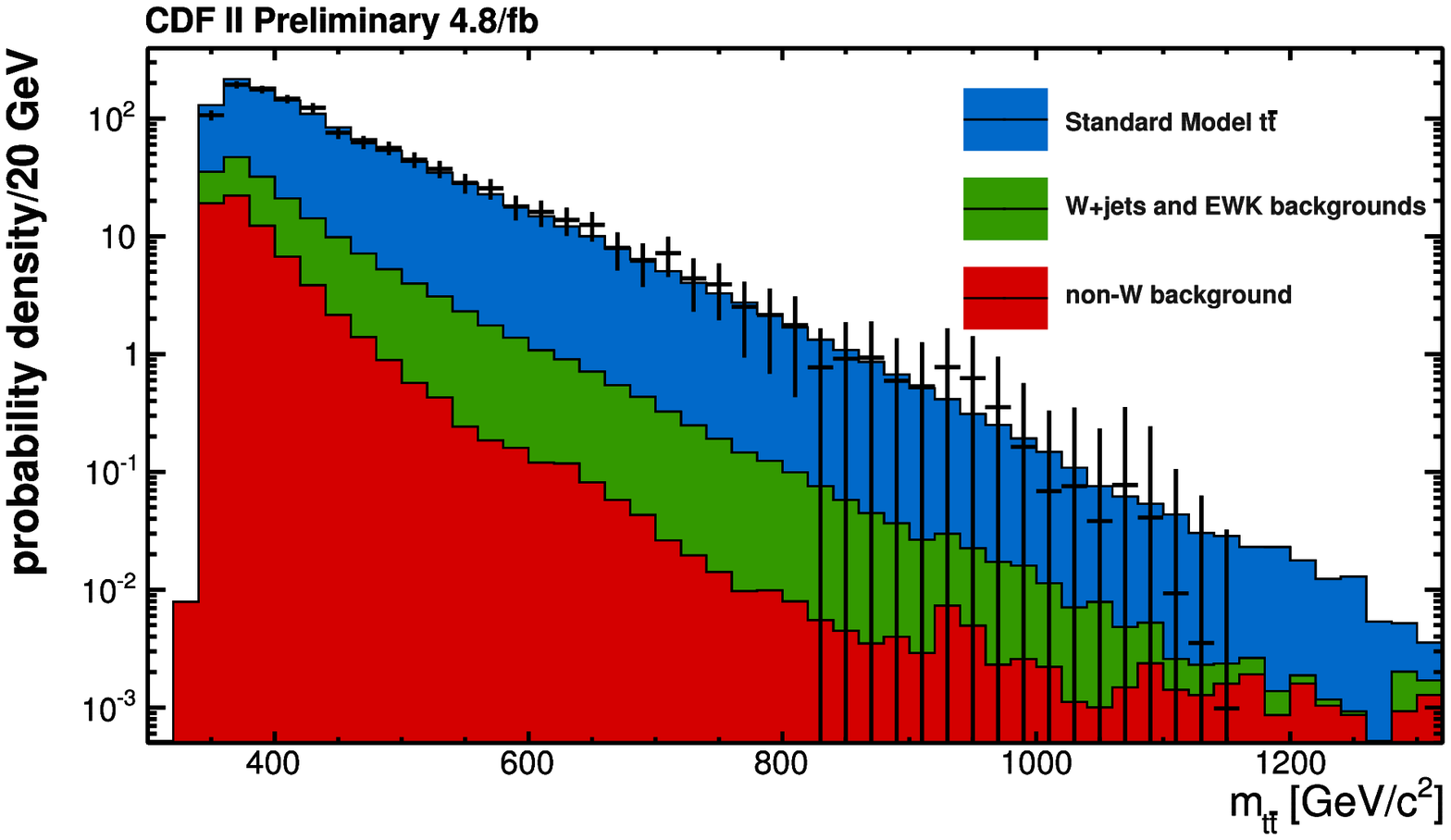}
\caption{
Unfolded $d\sigma/dM_{t\bar{t}}$ distribution compared to \pythia\ (left) and a
NLO+NNLL QCD calculation (middle). Right:
$ M_{t\bar{t}}$ distribution for SM \ttbar\ production and backgrounds
compared to data.
} \label{fig:cdf_mttbar}
\end{figure*}

The CDF Collaboration has measured the \ttbar\
differential cross section in the \ljets\ channel with respect to the \ttbar\ invariant mass,
$d\sigma/dM_{t\bar{t}}$, using an
integrated luminosity of 2.7~\fbmone~\cite{cdf_xsec_mttbar}.
The \ttbar\ invariant mass spectrum is a powerful test of higher order
QCD calculations and also sensitive to a variety
of exotic particles decaying into \ttbar\ pairs.
The \ttbar\ invariant mass is reconstructed, using the four-vectors of
the $b$-tagged jet and the three remaining leading jets in the event,
the lepton and the transverse components of the neutrino momentum,
given by missing transverse energy. The expected contribution from the backgrounds is
subtracted from the original $M_{t\bar{t}}$ distribution. The resulting $M_{t\bar{t}}$
signal distribution suffers from resolution smearing and is corrected
using a regularized unfolding technique, which also accounts for the
longitudinal component of the neutrino momentum. 
The measured $d\sigma/dM_{t\bar{t}}$ is shown in Fig.~\ref{fig:cdf_mttbar} (left).
The result is consistent with the SM expectation, as
modeled by \pythia\ with CTEQ5L parton distribution functions.
However, the description by \pythia\ is not perfect. 
A significantly better description of the data is achieved
in calculations where threshold logarithms
up to NNLL order are resummed~\cite{SMtheory_A}. Fig.~\ref{fig:cdf_mttbar} (middle)
shows a good agreement  in the $d\sigma/dM_{t\bar{t}}$ distribution with
the same CDF data.

\section{Searches for New Physics in Top Quark Pair Production}
The fact that the top quark has by far the largest mass of all known elementary
particles suggests that it may play a special role in
the dynamics of electroweak symmetry breaking. This makes searches for
new physics in the top quark sector very attractive. In the following
one recent example for a search for new physics in the
production of top quarks is presented: a search for \ttbar\
resonances which explores the invariant \ttbar\ mass distribution, too.

One of the various models incorporating the possibility of a special
role of the top quark in the dynamics of electroweak symmetry breaking
is topcolor~\cite{topcolor}, where the large top quark mass can be generated
through a dynamical \ttbar\ condensate, $X$, which is formed by a new strong
gauge force preferentially coupled to the third generation of
fermions. In one particular model, topcolor-assisted technicolor~\cite{Zprime},
$X$ couples weakly and symmetrically to the first and second generations
and strongly to the third generation of quarks, and has no couplings
to leptons, resulting in a predicted cross section for \ttbar\ production
larger than SM prediction. 

The CDF and D0 collaborations presented updated
model-independent searches for a narrow-width heavy resonance $X$
decaying into \ttbar\ using 
2.8~\fbmone\ of data in
the all-hadronic final 
state (CDF)~\cite{resonances_cdf}
and 
3.6~\fbmone\ of data analyzing $\ell$+jets
final states (D\O)~\cite{resonances_d0}.
A new search for resonant production of \ttbar\ is
performed by the CDF collaboration in the \ljets\ channel using a data
set of 4.8~\fbmone.
The search for resonant production is
performed by examining 
the reconstructed \ttbar\ invariant mass distribution which is shown
in Fig.~\ref{fig:cdf_mttbar} (right).
 A matrix element reconstruction technique is used
and for each event a probability density function (pdf) of the \ttbar\ invariant
mass is sampled. These pdfs are used to construct a likelihood
function, whereby the cross section for resonant \ttbar\ production,
given a hypothetical resonance mass and width, is estimated. 
One compares the 95\% C.L. points for these posterior probabilities
for the data with those expected from SM sources only and determines
how likely it is for the SM to fluctuate to the data for each mass
point. A very unlikely fluctuation could then indicate the presence of
new physics in the sample. 
There is no evidence of resonant production of \ttbar\ candidate
events. Therefore within a top-color-assisted technicolor model, the existence of a
leptophobic $Z'$ boson with $M_{Z'} < 900$~GeV and width around $\Gamma_{Z'}	=
0.012 M_{Z'}$ is excluded at 95\% CL.

\section{SINGLE TOP QUARK PRODUCTION}

\subsection{Single Top Quark Production Cross Section and $V_{tb}$}

Single top quark
production serves as a probe of the $Wtb$
interaction~\cite{singletop-wtb}, and its production cross section
provides a direct measurement of the magnitude of the quark mixing
matrix element $V_{tb}$ without assuming three quark
generations~\cite{singletop-vtb-jikia}. However, measuring the yield
of single top quarks is difficult because of the small production rate
and large backgrounds. At the Tevatron single top quarks are produced
by either a $t$-channel exchange of a virtual $W$ boson which combines with
a highly energetic $b$ quark to produce a top quark, or by an
$s$-channel exchange of a far off-shell $W$ boson which decays to
produce a top quark and a $b$ antiquark. 

In March 2009, the CDF and D0 Collaborations reported
observation of the electroweak production of single 
top quarks in {\ppbar} collisions at $\sqrt{s} = 1.96$~TeV based on
3.2~fb$^{-1}$ (CDF) ~\cite{sitop_obs_cdf,sitop_obs_cdf_prd} and 2.3~fb$^{-1}$ (D\O)
~\cite{sitop_obs_d0} of 
data. Both 
Collaborations used events containing an isolated electron or 
muon and missing transverse energy, together with jets where one or
two of the jets were required to originate 
from the fragmentation of $b$ quarks. The data
are devided into independent sets (subchannels) depending on lepton flavor, jet
multiplicity, number of $b$-tags and in the case of D0 the run period.

\begin{figure*}[t]
\centering
\includegraphics[width=170mm]{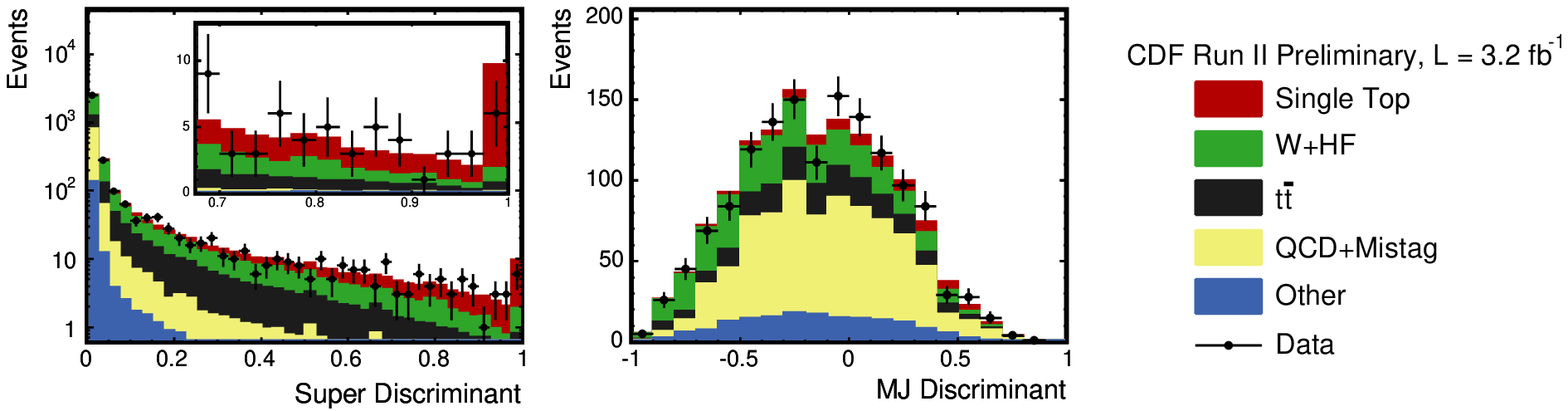}\\
\includegraphics[width=60mm]{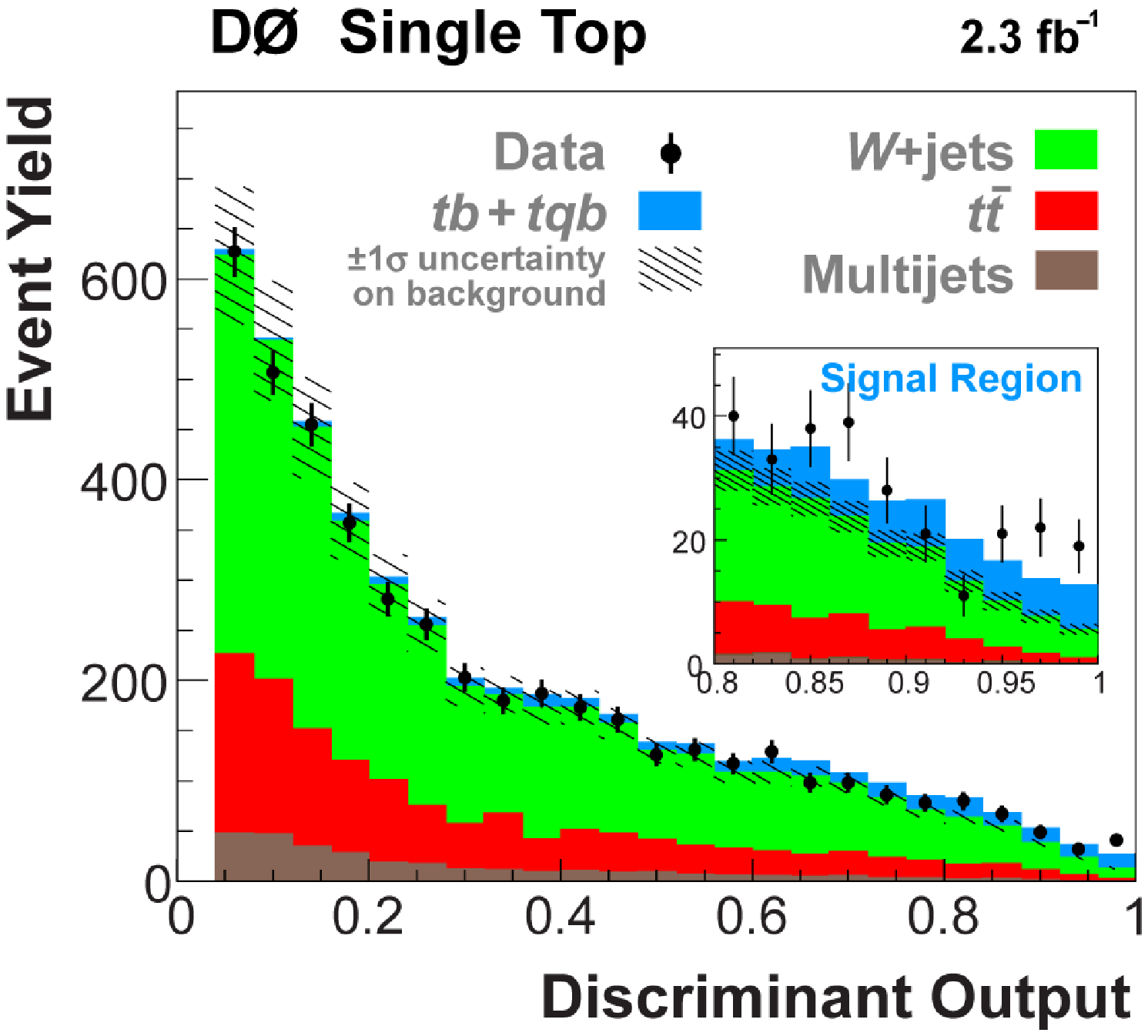}
\includegraphics[width=71mm]{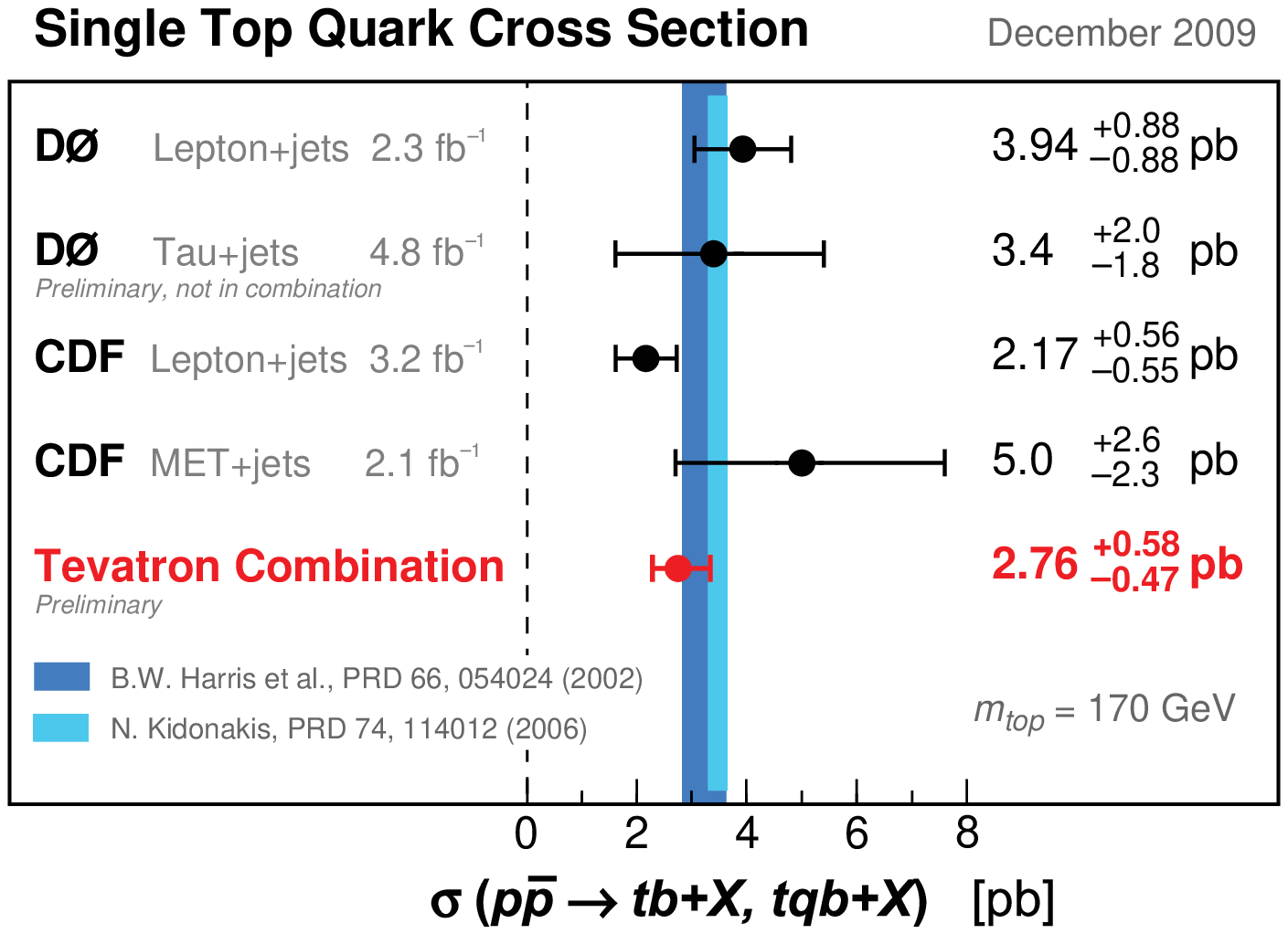}
\caption{Output discriminants for single top quark
  observation. Displayed are the CDF super-discriminant output (upper left),
  the CDF discriminant output for the analysis using a missing transverse
  momentum and jets selection (upper middle) and the D0 BNN
discriminant (lower left). A summary of all existing measurements and
  the Tevatron combination is shown (lower right).
} \label{fig:sitop}
\end{figure*}

CDF and D0 each combine many variables using different multivariate
analysis methods such as Boosted Decision Trees, Neural Networks,
Bayesian Neural Networks, Matrix Elements and Likelihood functions to
increase the separation power between signal and background. 
The discriminant outputs of each multivariate analysis are combined to one
discriminant taking the correlations into account.  CDF combines 8 subchannels into a
super-discriminant using a neural network trained with
neuroevolution~\cite{Nevolution} which is shown in
Fig.~\ref{fig:sitop} (upper left).
The CDF Collaboration has an additional independent search
channel~\cite{sitop_obs_cdf,sitop_mj_cdf}
that is designed to select events with missing transverse energy and
jets and is orthogonal to the other channels. It accepts events in
which the $W$ boson decays 
into hadronically decaying $\tau$ leptons and recovers events lost
due to electron or muon identification inefficiencies. The
discriminant output of this analysis -- shown in Fig.~\ref{fig:sitop}
(upper middle) -- is combined to the
super-discriminant to obtain the final result.  D0 combines 12
subchannels into one BNN discriminant displayed in
Fig.~\ref{fig:sitop} (lower left).

CDF measures a cross section of
$\sigma({\ppbar} \rightarrow\ tb+X,~tqb+X) =2.3 ^{+0.6}_{-0.5}$~pb assuming
a top quark mass of 175~GeV
and 
D0 measures a cross section of
$\sigma({\ppbar} \rightarrow\ tb+X,~tqb+X) = 3.94 \pm 0.88$~pb at a top
quark mass of 170~GeV.
Both measurements correspond to a
5.0~standard deviation (SD) significance for the observation. They are in
agreement with the SM predictions.
Both results are translated into a direct measurement
of the amplitude of the CKM matrix element $V_{tb}$ without making
assumptions on the number of quark generations and the matrix unitarity.
CDF obtains $|V_{tb}| = 0.91 \pm 0.13$, D0 derives $|V_{tb}| = 1.07 \pm 0.12$.

All measurements mentioned
above~\cite{sitop_obs_cdf,sitop_obs_d0,sitop_mj_cdf} and a measurement
by the D0 Collaboration analyzing the $\tau$+jets final
state by explicitly identifying hadronically decaying $\tau$
leptons~\cite{sitop_tau_d0}, 
agree with the SM predictions~\cite{SMsitop_H,SMsitop_K}.
This can be seen in Fig.~\ref{fig:sitop} (lower right). The Tevatron
combination (still excluding the $\tau$+jets final
state)~\cite{sitop_combi} gives a single top cross section of
\begin{eqnarray}
\sigma({\ppbar} \rightarrow\ tb+X,~tqb+X) = 2.76^{+0.58}_{-0.47}\:{\rm pb} \nonumber
\end{eqnarray}
and a CKM matrix element of
\begin{eqnarray}
|V_{tb}| = 0.88 \pm 0.07\,. \nonumber
\end{eqnarray}

\subsection{$s$-Channel and $t$-Channel Single Top Quark Production}

In the observation analyses the combined $s+t$ channel
single top 
quark cross section was measured, assuming the SM ratio of the two production
modes. This ratio is modified in several new physics scenarios,
for example in models with additional quark generations, new heavy
bosons, flavor-changing neutral currents, or anomalous top quark
couplings. Therefore it is interesting to remove this constraint and
to use the $t$-channel 
characteristics to measure the $t$-channel and $s$-channel
cross sections simultaneously which provides a $t$-channel
measurement independent of the $s$-channel cross section model.  

The D0 Collaboration reported direct evidence for the
electroweak production of single top quarks through the $t$-channel
exchange of a virtual $W$ boson alone~\cite{tchannel_d0}. The measured
cross section is $\sigma(t-{\rm channel}) = 
3.14^{+0.94}_{-0.81}$~pb, has a significance of 4.8 SD and is
consistent with the SM prediction.  
This is the first analysis to isolate an
individual single top quark production channel. 
Both the CDF and the D0 Collaborations extracted the $s$-channel
($\sigma_s$) and $t$-channel ($\sigma_t$) cross sections
simultaneously relaxing the SM ratio assumption for
$\sigma_s$/$\sigma_t$~\cite{sitop_obs_cdf_prd,tchannel_d0}. The
results agree with the SM predictions. 

\subsection{Searches for New Physics in Single Top Quark Production}
\begin{figure*}[t]
\centering
\includegraphics[width=60mm]{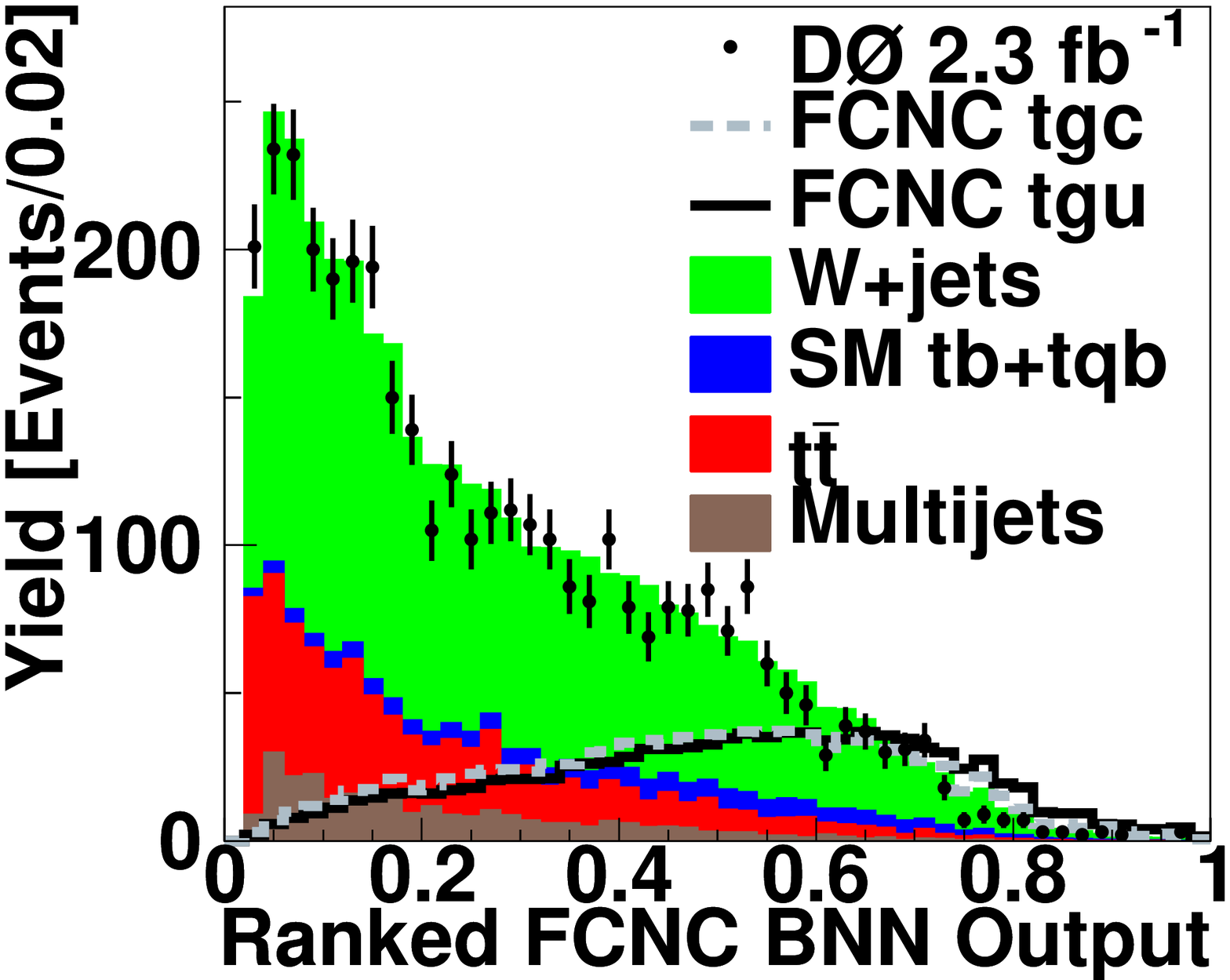}
\includegraphics[width=53mm]{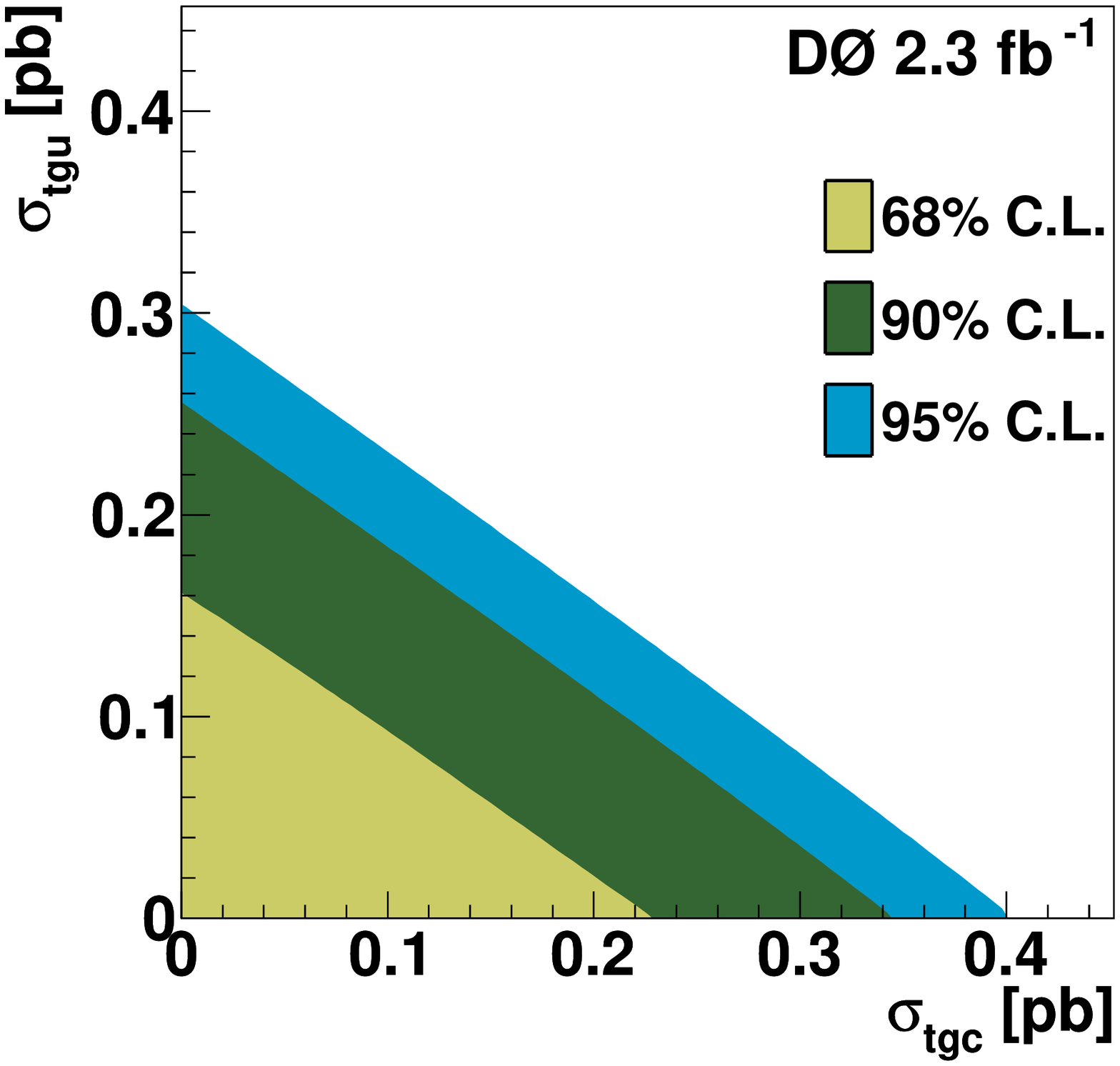}
\caption{Left: comparison of the background model to data for the FCNC
  discriminant summed over all analysis channels. The bins have
  been ordered by their signal to background ratio and the FCNC
  signals are each normalized to a cross section of 5~pb. Right:
  Bayesian posterior probability as a function of the $\sigma_{tgu}$ and
  $\sigma_{tgc}$ cross sections. 
} \label{fig:d0_sitop_fcnc}
\end{figure*}

The single top quark production channel offers a large variety for
sensitive searches for new physics beyond the SM. 
One example are searches for flavor
changing neutral current (FCNC) couplings~\cite{cdf_FCNC,d0_FCNC}.
Recently, the D0 Collaboration searched
for single top quark production via 
FCNC top-gluon-quark couplings $\kappa_{tgu}$ and $\kappa_{tgc}$ in
a sample corresponding to 2.3~\fbmone\
of integrated luminosity collected~\cite{d0_FCNC}. Events containing a single top
quark candidate with an additional jet are selected. Separation
between signal and background is obtained using Bayesian neural
networks (BNN) which are trained for each choice of lepton
flavor (electron or muon), jet multiplicity (2, 3, or 4), and
data-taking period separately, twelve in total.
 Fig.~\ref{fig:d0_sitop_fcnc} (left) shows the comparison between
 background and data for all 
 twelve BNN discriminants combined. 
Since the data are consistent with the background expectation, upper
limits on the FCNC cross sections and couplings are set using a
Bayesian approach~\cite{bayesian}. Following the analysis strategy of previous
work~\cite{post_density}, a two-dimensional Bayesian posterior density for the cross
sections and for the square of the FCNC couplings is formed, using the
BNN distributions for data, background, and signals. 
Fig.~\ref{fig:d0_sitop_fcnc} (right) shows the posterior density as a
function of the FCNC cross sections $\sigma_{tgu}$ and
$\sigma_{tgc}$. 
Since there is agreement with the SM, 
limits are set on the couplings of $\kappa_{tgu}/\Lambda < 0.013$
TeV$^{-1}$ and $\kappa_{tgc}/\Lambda < 0.057$ TeV$^{-1}$, $\Lambda$
being the scale of the new interactions which generate these couplings (of order 1 TeV),
without making assumptions about the $tgc$ and $tgu$ couplings, respectively. 

\section{Conclusions}

Recent highlights in top quark production analyses from the Tevatron
collider are reviewed. Among many impressive results, the 
observation of single top quark production via the electroweak interaction
and the direct measurement of the CKM matrix element $V_{tb}$ by the
CDF and D0 Collaborations is
outstanding. 

Analyzing datasets corresponding to an integrated
luminosity of up to 5.7~\fbmone\ the Tevatron experiments can
perform high precision measurements. As an example the top quark
pair production cross section is extracted with an accuracy of
6\%. The high statistics in \ttbar\ samples allows to analyze
differential cross sections for the first time with a high 
accuracy resulting in powerful tests of higher order QCD
calculations. In general the inclusion of soft gluon resummations at the NNLL
level into NLO QCD calculations improves the description of the
data. Both top quark pair 
production and single top quark 
production channels allow sensitive searches for new phenomena beyond
the SM. So far, there is no hint for new physics beyond the SM in the
top sector. Excellent prospects for top quark physics investigations can be
expected with more integrated luminosity at the Tevatron.

%

\end{document}